%% file: 8_main.tex
\documentclass[11pt]{article}

\input{0_usepackage}

\title{\ours: Unified Optical Vision-Language Representation for Multimodal Recommendation}

\author{
\textbf{Yueqi Wang}\textsuperscript{1}\thanks{Equal contribution.} \quad
\textbf{Zitian Guo}\textsuperscript{1}\footnotemark[1] \quad
\textbf{Yupeng Hou}\textsuperscript{1} \quad
\textbf{Yifei Wang}\textsuperscript{2} \quad
\textbf{Kibum Kim}\textsuperscript{1} \quad
\textbf{Zhenrui Yue}\textsuperscript{3} \\
\textbf{Shuo Xing}\textsuperscript{6} \quad
\textbf{Haodong Li}\textsuperscript{1} \quad
\textbf{Heming Xia}\textsuperscript{4} \quad
\textbf{Renrui Zhang}\textsuperscript{5} \quad
\textbf{Zhengzhong Tu}\textsuperscript{6} \quad
\textbf{Julian McAuley}\textsuperscript{1} \\
\textsuperscript{1}University of California, San Diego \quad
\textsuperscript{2}Alibaba Group \quad
\textsuperscript{3}University of Illinois Urbana-Champaign \\
\textsuperscript{4}The Hong Kong Polytechnic University \quad
\textsuperscript{5}The Chinese University of Hong Kong \quad
\textsuperscript{6}Texas A\&M University
}

\begin{document}
\maketitle
\begin{abstract}
Recent advances in vision-language modeling have substantially improved multimodal encoding, retrieval and reasoning. Yet for multimodal recommendation, encoding rich item vision-language semantic interactions remains a long-standing bottleneck, which hampers accurate item representation learning and user-item matching. Mainstream approaches primarily adopt independent encoding of vision and language modality followed by rigid late fusion such as concatenation, inherently omitting native vision-language interactions and introducing cross-modal semantic distortion. To address this challenge, we propose \ours, the first visual-space unified encoding paradigm for multimodal collaborative filtering, a fundamental recommendation setting. Instead of isolated modality-specific encoding, \ours renders item textual metadata as visual glyphs, enabling native image–text interaction within visual encoder---the \textit{perceptual encoding} level. The resulting representations are further processed by the language decoder---the \textit{semantic encoding} level, allowing \ours to exploit the \textit{dual-attention} mechanism of modern vision-language models that previous encoding methods omitted. \ours's efficacy is theoretically supported by mutual information analysis and empirically demonstrated through superior performance across strong baselines and benchmarks. As a plug-and-play module, \ours (1) introduces minimal cost, (2) is robust against rendered text font, color and layout, etc., and (3) integrates seamlessly into existing multimodal collaborative filtering models.
\end{abstract}

\input{1_intro}
\input{2_related}
\input{3_method}
\input{4_setup}

\input{5_exp}
\input{6_ablation}

\section*{Limitations}
This work studies unified optical encoding under feature-based multimodal collaborative filtering, which is a mainstream setting for large-scale item encoding. Although we evaluate multiple backbones, rendering settings, text lengths, and VLM scales, broader validation across additional recommendation domains such as video recommendation remains valuable for future research.


\bibliography{custom}

\input{7_appendix}

\end{document}

%% file: 0_usepackage.tex
\usepackage[preprint]{acl}

\usepackage{times}
\usepackage{latexsym}

\usepackage[T1]{fontenc}

\usepackage[utf8]{inputenc}

\usepackage{microtype}

\usepackage{inconsolata}

\usepackage{graphicx}

\usepackage{booktabs}
\usepackage{multirow}

\usepackage{enumitem}
\usepackage{url}
\usepackage{adjustbox}
\usepackage{subcaption}
\usepackage{amsthm}
\newtheorem{property}{Property}
\usepackage{amsmath}
\usepackage{amssymb}
\usepackage{cleveref}

\usepackage{xspace}
\newcommand{\ours}{OpticalRec\xspace}

%% file: 1_intro.tex
\section{Introduction}
\label{sec: intro}
Recent progress in vision-language modeling and multimodal learning has opened new opportunities for recommender systems~\cite{singh2026openaigpt5card, li2024llavaonevision, bai2025qwen3vltechnicalreport, xing2025re, kim2026sink}. These improvements mainly stem from two aspects: First, textual and visual representations offer rich semantic signals that help characterize user interests and item properties. Second, the above representations learned during large-scale vision-language pretraining are transferrable to recommendation tasks~\cite{malitesta2025formalizing, xu2025survey}. As a result, large vision-language model representations have become a powerful basis for improving recommendation quality, contextual understanding, and personalization in modern recommender systems~\cite{pomo2025recommender, ye2025multimodal}.

Despite above mentioned benefits of large vision-language models, injecting rich multimodal features into recommender systems continues to pose significant challenges. A fundamental bottleneck lies in the widely-adopted decoupled/independent encoding of each modality followed by the lossy late fusion~\cite{xu2025survey, malitesta2025formalizing}. Two major disadvantages exist here: 1. independent encoding inherently omits vision-language cross-interactions such as cross attention and yields embeddings in heterogeneous vector space per-modality, 2. the lossy late fusion such as summation, concatenation and graph fusion, is applied to integrate the heterogeneous embedding in 1., which distorts the cross-modal semantics in resulting encodings. Specifically, summation or concatenation-based late fusion omits vision-language cross interactions with fusion-entailed information loss~\cite{zhou2025doesmultimodalityimproverecommender}; Graph-based methods exploit multimodal fusion by similarity graphs, but initial graph construction still falls back to late-fused separate encoding~\cite{zhang2021mining, zhou2023tale}. Recent works continue to refine downstream graph architectures while overlooking image-text interactions and the disruptive late fusion~\cite{ong2025spectrum, yu2025mind}. Hence, we analyze that a unified-space encoding paradigm is optimal for multimodal recommendation that satisfies:
\begin{enumerate}
    \item Multimodal content should be presented in a unified input space, allowing rich vision-language interactions instead of isolation.
    \item The unified input should be encoded once into a homogeneous embedding space, without additional disruptive late fusion.
\end{enumerate}

Motivated by these observations~\cite{bai2025qwen3vltechnicalreport, wang2025internvl35advancingopensourcemultimodal}, we propose \ours, a lightweight, plug-and-play unified encoding framework for multimodal recommendation. Through an efficient, one-time rendering operation that transforms textual metadata into visual text/glyph, \ours enables rich images and text cross-interactions within a unified pixel space during vision encoding. This also naturally aligns with a “see-as-users-see” principle: users shift attention back-and-forth between item image and text when browsing online shopping content. Further processed by the language decoder, \ours together forms a \textit{dual attention} mechanism and outputs homogeneous embeddings without late fusion. We summarize our contribution as follows:

\begin{enumerate}
    \item We propose \ours, the first visual-space unified encoding framework for multimodal collaborative filtering allowing rich, \textit{dual attention}-based image-text cross-interactions.
    \item \ours preserves vision-language correlations without the widely adopted, lossy late fusion, and is supported by theoretical analysis via mutual information and matrix effective rank.
    \item Experiments across multiple Amazon Review benchmarks and recommendation backbones demonstrate \ours's consistent improvements over strong multimodal baseline encoders.
    \item \ours is lightweight, plug-and-play, and robust against input layout, text color, font etc., making it readily adaptable to existing multimodal collaborative filtering frameworks.
\end{enumerate}

%% file: 2_related.tex
\section{Related Works}
\label{sec: related}
\subsection{Multimodal Collaborative Filtering}
Pretrained language and vision-language models are widely adopted in multimodal collaborative filtering (MMCF) for transfer learning~\cite{malitesta2024we, xu2025survey}. A major line of work adopts pretrained multimodal embeddings as frozen item features for collaborative filtering~\cite{xin2019relational, wang2021dualgnn, zhou2025learning, li2024neighborhood, sheng2024language}. Early methods such as VBPR~\cite{he2016vbpr} leverage CNN-based image features to model visually driven user preferences. Later methods further introduce textual features and graph neural networks for high-order multimodal interaction modeling~\cite{chen2017attentive, zhou2023mmrec, yu2023multi, liu2023megcf, zhou2023enhancing}. For example, MMGCN~\cite{wei2019mmgcn} and GRCN~\cite{wei2020graph} extend user-item graph with modality-aware fusion. Later developments such as LATTICE~\cite{zhang2021mining} and FREEDOM~\cite{zhou2023tale} introduce item-item content similarity graphs for augmentation. Recent methods continue to explore tailored graph structures and better training supervision~\cite{zheng2024enhancing, xu2025survey, liu2025modalsync, guo2024lgmrec, zhou2023bootstrap}. In contrast, our work shifts the focus from increasingly specialized graph networks to unified encoding for more advanced image-text interactions.

Beyond feature-based MMCF, recent and past studies also explore multimodal sequential/generative recommendation and model behavior study~\cite{wang2024train, yue2024linear, wang2025mllm4rec, wang2025your, yue2025preference, zhai2025multimodal, lei2026unirec, qiao2026text}. We focus on MMCF for a cleaner encoding setup, where item representations directly interact with collaborative signals.

\subsection{Vision-Language Grounding and Unified Encoding}
Feature-based multimodal recommendation predominantly follows a separate-encoding paradigm plus late fusion~\cite{ye2025multimodal, he2025understanding}. This paradigm aligns with early vision-language representation learning such as CLIP, which learns shared image-text embeddings but still encodes each modality separately. Later models such as Flamingo and LLaVA move toward joint, interleaved vision-language modeling. More recent models such as Qwen-VL~\cite{bai2025qwen3vltechnicalreport}, InternVL~\cite{wang2025internvl35advancingopensourcemultimodal}, and LLaVA model series~\cite{li2024llavaonevision, li2024llavanextinterleave} further strengthen visual grounding and layout-aware understanding. However, how such unified vision-language perception can benefit MMCF remains unexplored, as user-item collaborative signal is inherently multifaceted. We aim to address this challenge in the following of this work.

%% file: 3_method.tex
\input{materials/fig_code_main}
\section{Methodologies}
\subsection{Multimodal Collaborative Filtering Recommendation}
\label{subsec: multimodal cf definition}
We briefly introduce the Multimodal Collaborative Filtering (MMCF) task. Specifically, Let $\mathcal{U} = \{u_1, u_2, \ldots, u_M\}$ and $\mathcal{I} = \{i_1, i_2, \ldots, i_N\}$ denote the sets of $M$ users and $N$ items. An item $i$ is described by an image $\mathbf{v}_i$ and textual description $\mathbf{t}_i$, item embedding $\mathbf{q}_i$ is obtained by encoder $f$:

\begin{equation}
    \mathbf{q}_i = f(\mathbf{v}_i, \mathbf{t}_i),~~~\mathbf{p}_u = \mathcal{A}(\{\mathbf{q}_i\}_{i \in \mathcal{N}_u})
\end{equation}
where $\mathcal{N}_u$ is item set user $u$ interacted and $\mathcal{A}(\cdot)$ is the aggregation function. Recommender $\mathcal{F}(\cdot)$ further transforms $\mathbf{p}_u$, $\mathbf{q}_i$, and user-item graph $\mathcal{G}$: $(\mathbf{p}_u', \mathbf{q}_i') = \mathcal{F}(\mathbf{p}_u, \mathbf{q}_i, \mathcal{G})$. Finally, the likelihood that item $i$ is recommended to user $u$ is defined as their inner-product:
\begin{equation}
y_{ui} = \langle \mathbf{p}_u', \mathbf{q}_i' \rangle
\end{equation}

 
\subsection{\ours as Pixel-Space Unified Encoding}
\label{subsec: OpticalRec definition}
As discussed in \Cref{sec: intro}, commonly adopted independent encoding with late fusion suffers from 1. lack of vision-language interaction and 2. rigid fusion corruption. We therefore propose \ours, the first unified pixel-space encoding for multimodal recommendation with advanced cross-modal interactions and is late-fusion free.

Specifically, both independent encoding + late fusion and \ours can be expressed as Markov chains as current output depends solely on the previous step.

Independent encoding is shown in \Cref{fig: main figure}(a), which processes each modality separately with vision and language encoders $f_V, f_T$ into modality-specific vectors $(\mathbf{z}_i^v, \mathbf{z}_i^t)$ before applying a fusion operator $g$:
\begin{equation}
    (\mathbf{v}_i, \mathbf{t}_i)
    \;\xrightarrow{f_V,\, f_T}\;
    \mathbf{z}_i^{\text{LF}} \triangleq (\mathbf{z}_i^v, \mathbf{z}_i^t)
    \;\xrightarrow{g}\;
    \mathbf{q}_i^{\text{LF}}
\end{equation}
where $\mathbf{z}_i^{\text{LF}}$ denotes output of two independent encoders requiring late fusion and $\mathbf{q}_i^{\text{LF}}$ means late-fused final embedding. Prompt-based encoding, shown in \Cref{fig: main figure}(b), improves the independent encoding by introducing a decoder-level vision-language interaction. However, the image and text are still encoded independently, as projected image tokens are concatenated together with text tokens, which is a late-fusion operator $g(\cdot)$, before feeding into the language decoder.

\ours as unified encoding first applies an efficient rendering operation $\mathcal{R}(\cdot)$ that transforms into a homogeneous pixel space $\mathbb{R}^d$ all optical elements such as (1) item image $\mathbf{v}_i$, (2) item textual description $\mathbf{t}_i$, and (3) image-text layout structure $\mathbf{s}^{vt}_{i}$ following visual organization of product pages:
\begin{equation}
    \tilde{\mathbf{v}}_i = \mathcal{R}(\mathbf{v}_i, \mathbf{t}_i, \mathbf{s}^{vt}_{i})
\end{equation}
The rendering operation is performed once offline and introduces negligible overhead discussed \Cref{sec: ablation}. Specifically, we perform a simple rendering of the item textual-meta data into the pixel space together with the item image in the following steps:
\begin{enumerate}[label=(\arabic*), leftmargin=*]
    \item Create a blank canvas of square resolution $m \times m$ pixels, and set an image height value as $h$ $(h<m)$ pixels.
    \item Fit the original item image by setting its height as $h$, and adjust its width $w$ by maintaining the original image aspect ratio.
    \item Retrieve the text meta-data of \{Title, Brand, Category, Price\} by item id and render it into the canvas in the remaining $m \times (m-h)$ space. The layout and color should loosely follow the amazon website style.
\end{enumerate}

A rendered item card example is shown in top left of \Cref{fig: main figure}. Then, the unified input $\tilde{\mathbf{v}}_i$ is mapped to the final embedding $\mathbf{z}_i^{\text{OR}} = \mathbf{q}_i^{\text{OR}}$ using \textit{a single vision-language encoder $f_\Phi$} corresponding to \Cref{fig: main figure} (c) frozen VLM. The whole \ours encoding is illustrated as below and in \Cref{fig: main figure}(c):
\begin{equation}
  (\mathbf{v}_i, \mathbf{t}_i, \mathbf{s}^{vt}_{i})
  \;\xrightarrow{\mathcal{R}}\;
  \tilde{\mathbf{v}}_i
  \;\xrightarrow{f_\Phi}\;
  \mathbf{z}_i^{\text{OR}}
  \;=\;
  \mathbf{q}_i^{\text{OR}}
\end{equation}
Consequently, \ours produces a compressed, unified item representation with \textit{dual-attentive} vision-language interactions in the vision encoder---the perception level and in the language decoder level---the semantic-level~\cite{bai2025qwen3vltechnicalreport, wang2025internvl35advancingopensourcemultimodal}. The perception level interaction happens in the vision encoder introduces high-level image-text regional matching, which has never been explored in MMCF encoding as we further analyze in \Cref{sec: Experimental Results}.

\subsection{Mutual Information Gap between the Unified and Late-Fusion Encoding}
\label{subsec: mutual information gap}
In the previous section, we introduced how \ours encodes multimodal content. In this section, we analyze encoded representation informativeness of \ours and late-fusion-based encoding, and demonstrate why \ours is theoretically more expressive with an even greater advantage in practical-scale settings.

Specifically, we compare the information differences by introducing the mutual information gap. Let $\mathcal{I}_{\text{cross}}^{\text{OR}}$
denote the cross-modal interaction captured by \ours's, let $\mathcal{D}_{\text{intra}}^{\text{LF}}$
denote the information loss during late fusion $g$. The information gap then decomposes as:
\begin{equation}
\resizebox{0.89\linewidth}{!}{$
  I\!\left(\mathbf{q}_i^{\text{OR}};\, (\mathbf{v}_i, \mathbf{t}_i)\right)
  \;-\;
  I\!\left(\mathbf{q}_i^{\text{LF}};\, (\mathbf{v}_i, \mathbf{t}_i)\right)
  \;=\;
  \mathcal{I}_{\text{cross}}^{\text{OR}}
  \;+\;
  \mathcal{D}_{\text{intra}}^{\text{LF}}
  \label{eq:gap}
$}
\end{equation}
Here, $I(X;Y)$ denotes mutual information, measuring how much information X preserves about Y. Intuitively, \Cref{eq:gap} compares the information retained by \ours and late-fusion about the original image-text input; we next derive why both right-hand terms are positive and that their sum is substantial in practice.

\paragraph{Brief positivity proof}
Let $\mathbf{x}_i=(\mathbf{v}_i,\mathbf{t}_i)$ denote the image--text input, and let $\mathbf{q}_i^{\text{OR}}$ be the \ours embedding. We show that both terms in \Cref{eq:gap} are positive; the full proof is provided in Appendix~\ref{app:mutual_information_gap}.

The cross-modal information gain of unified encoding is
\begin{equation}
\label{eq:cross_modal_information_gain}
\mathcal{I}_{\text{cross}}^{\text{OR}}
\triangleq
I(\mathbf{q}_i^{\text{OR}};\mathbf{x}_i)
-
I(\mathbf{z}_i^{\text{LF}};\mathbf{x}_i).
\end{equation}
Independent encoding cannot fully preserve cross-modal information, whereas \ours exposes both modalities to a unified vision-language encoder. Therefore,
$I(\mathbf{q}_i^{\text{OR}};\mathbf{x}_i)
>
I(\mathbf{z}_i^{\text{LF}};\mathbf{x}_i)$,
implying $\mathcal{I}_{\text{cross}}^{\text{OR}}>0$.

The information loss introduced by late fusion is
\begin{equation}
\label{eq:late_fusion_information_loss}
\mathcal{D}_{\text{intra}}^{\text{LF}}
\triangleq
I(\mathbf{z}_i^{\text{LF}};\mathbf{x}_i)
-
I(\mathbf{q}_i^{\text{LF}};\mathbf{x}_i).
\end{equation}
Since $\mathbf{q}_i^{\text{LF}}=g(\mathbf{z}_i^{\text{LF}})$, the data processing inequality~\cite{cover2006} gives
$I(\mathbf{q}_i^{\text{LF}};\mathbf{x}_i)
\leq
I(\mathbf{z}_i^{\text{LF}};\mathbf{x}_i)$.
For lossy fusion, the inequality is strict, yielding
$\mathcal{D}_{\text{intra}}^{\text{LF}}>0$.
We also provide an empirical significance analysis in Appendix~\ref{app: Information Gap Practical Implications}.

\subsection{Theoretical Analysis via Matrix Effective Rank}
\label{subsec: method matrix effective rank}
We use matrix effective rank $\operatorname{erank}(\cdot)$~\cite{roy2007effective}, a more sensitive alternative to true rank that measures the number of effectively utilized singular directions. For $\mathbf{X}\in\mathbb{R}^{m\times n}$, it is defined as
\begin{equation}
\operatorname{erank}(\mathbf{X})
=
\exp\left(-\sum_{i=1}^{n} p_i \log p_i\right),
\label{eq: erank of p i}
\end{equation}
where $p_i=\sigma_i/\sum_{j=1}^{n}\sigma_j$ is the normalized singular-value distribution. Lower values indicate representation collapse, whereas higher values reflect richer information spread. The derivation is provided in Appendix~\ref{app:effective_rank_derivation}, and the results are reported in \Cref{subsec: matrix effective rank results analysis}.

%% file: materials/fig_code_main.tex
\begin{figure*}[t!]
    \centering
    \resizebox{\textwidth}{!}{%
        \includegraphics{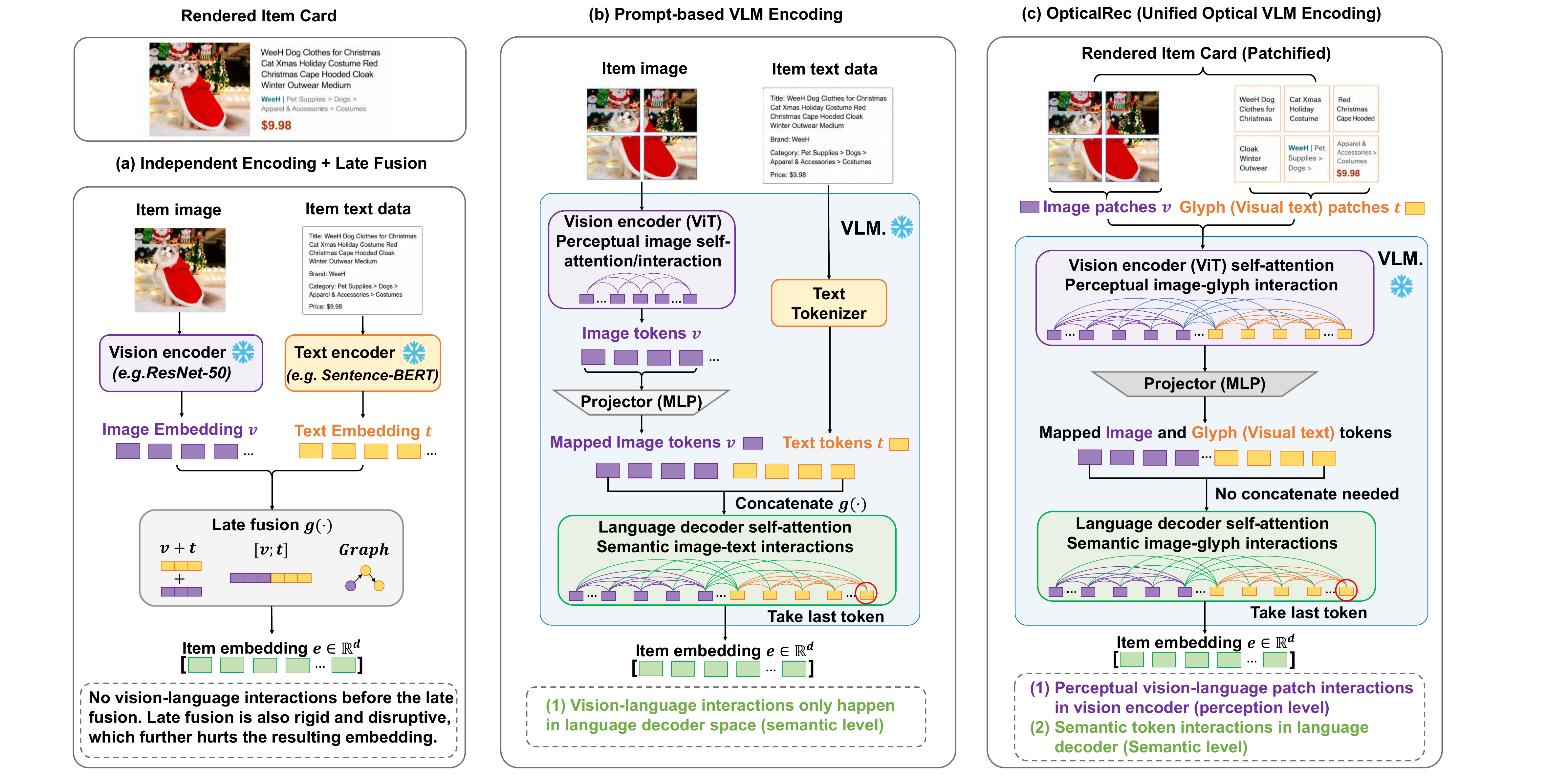}
    }
    \caption{Comparison of (a) conventional independent encoding with late fusion, (b) prompt-based VLM encoding, and (c) \ours. (a) Independently encodes image and text, allowing no cross-modal interaction before a lossy and rigid late-fusion step. (b) Also encodes the two modalities separately, but introduces semantic-level interaction in the language decoder. (c) \ours jointly encodes image and text in a unified visual space, enabling both 1. patch/perception-level interactions in the vision encoder and 2. mapped-token/semantic-level interactions in the language decoder. We regard the entire vision language model of (vision encoder, projector, language decoder) as an integrated VLM encoder, following mainstream architectures such as QwenVL and InternVL, etc. All encoders and VLMs are frozen/training free for computational efficiency.}
    \label{fig: main figure}
\end{figure*}

%% file: 4_setup.tex
\section{Experimental Setup}

\begin{table}[t]
\centering
\caption{Statistics of the datasets.}
\label{tab:dataset_stats}
\resizebox{\columnwidth}{!}{%
\begin{tabular}{lcccc}
\toprule
\textbf{Datasets} & \textbf{$|\mathcal{U}|$} & \textbf{$|\mathcal{I}|$} & \textbf{$|\mathcal{R}|$} & \textbf{Sparsity (\%)} \\
\midrule
\textbf{Baby}     & 108,897 & 17,836 & 330,771   & 99.983\% \\
\textbf{Pets}     & 461,828 & 61,060 & 1,772,584 & 99.993\% \\
\textbf{Clothing} & 398,796 & 75,616 & 2,109,975 & 99.993\% \\
\bottomrule
\end{tabular}
}
\end{table}

\subsection{Datasets}
For evaluation, we select three most commonly used MMCF benchmarks from \emph{Amazon.com} known for real-world sparsity, namely \emph{Baby Products} (Baby), \emph{Pet Supplies} (Pets), and \emph{Clothing, Shoes \& Jewelry} (Clothing)~\cite{hou2024bridging}. For preprocessing, apply 5-core and absent feature filtering detailed in Appendix~\ref{app: dataset details} due to space limit. Dataset statistics are reported in \Cref{tab:dataset_stats}.

\subsection{Baseline Methods}
\label{subsec: baseline methods}
We compare five representative multimodal recommendation baselines. \textit{VBPR}~\cite{he2016vbpr} incorporates CNN-based visual features into BPR, while \textit{LATTICE}~\cite{zhang2021mining} learns modality-specific item graphs and \textit{FREEDOM}~\cite{zhou2023tale} improves graph robustness through graph freezing and denoising. \textit{BM3}~\cite{zhou2023bootstrap} adopts negative-free self-supervised learning, whereas \textit{LGMRec}~\cite{guo2024lgmrec} models local and global multimodal correlations with hypergraphs. We additionally construct a strong prompt-based VLM baseline (\Cref{fig: main figure}(b)) to isolate \ours's perception-level interaction beyond decoder-level semantic interaction. Details are explained in Appendix~\ref{app: baseline methods}.

\subsection{Implementations}
We use Qwen3-VL-8B for the main experiments and keep all VLMs frozen during encoding for efficiency and fair comparison. All recommendation models are trained with AdamW and early stopping, with key hyperparameters tuned on the validation set; performance is evaluated using NDCG@20, Recall@20, and Hit Ratio@20. Details can be viewed in Appendix~\ref{app: Implementations}.

%% file: 5_exp.tex
\section{Experimental Results}
\label{sec: Experimental Results}
\input{materials/tab_main_v4}
\subsection{Main Performance Analysis}
We report in \Cref{tab:main} experimental results of independent encoding + late fusion, prompt-based encoding, and \ours encoding. Sum and Concat are standard late fusion methods; Graph denotes the graph neural network fusion module in their respective recommendation models, we refer readers to \Cref{subsec: baseline methods} for original papers of modeling details. \ours consistently outperforms the strongest baseline across all backbones and datasets. (1) It yields an average gain of 2.12\% across all models, metrics, and datasets, with the largest improvements on Pets (3.21\% on average; up to 5.82\% on NDCG@20 with FREEDOM), while gains on Baby and Clothing remain consistent at 1.65\% and 1.51\%. (2) Improvements are larger on NDCG@20 than Recall@20 (2.41\% vs.\ 1.97\%), indicating better top-ranked ordering in addition to stronger retrieval. (3) Graph-based backbones benefit more than VBPR, with average gains of 2.29\%, 2.88\%, and 2.29\% for LATTICE, FREEDOM, and LGMRec, respectively, versus 1.09\% for VBPR, suggesting that richer multimodal item representations further enhance semantic propagation over interaction graphs. To highlight the plug-and-play attribute of \ours, we also report results using InternVL3.5-8B as the encoding vision-language model, another performant VLM in Appendix~\ref{app: internvl results}.

\begin{figure}[t]
    \centering
    \includegraphics[width=0.90\linewidth]{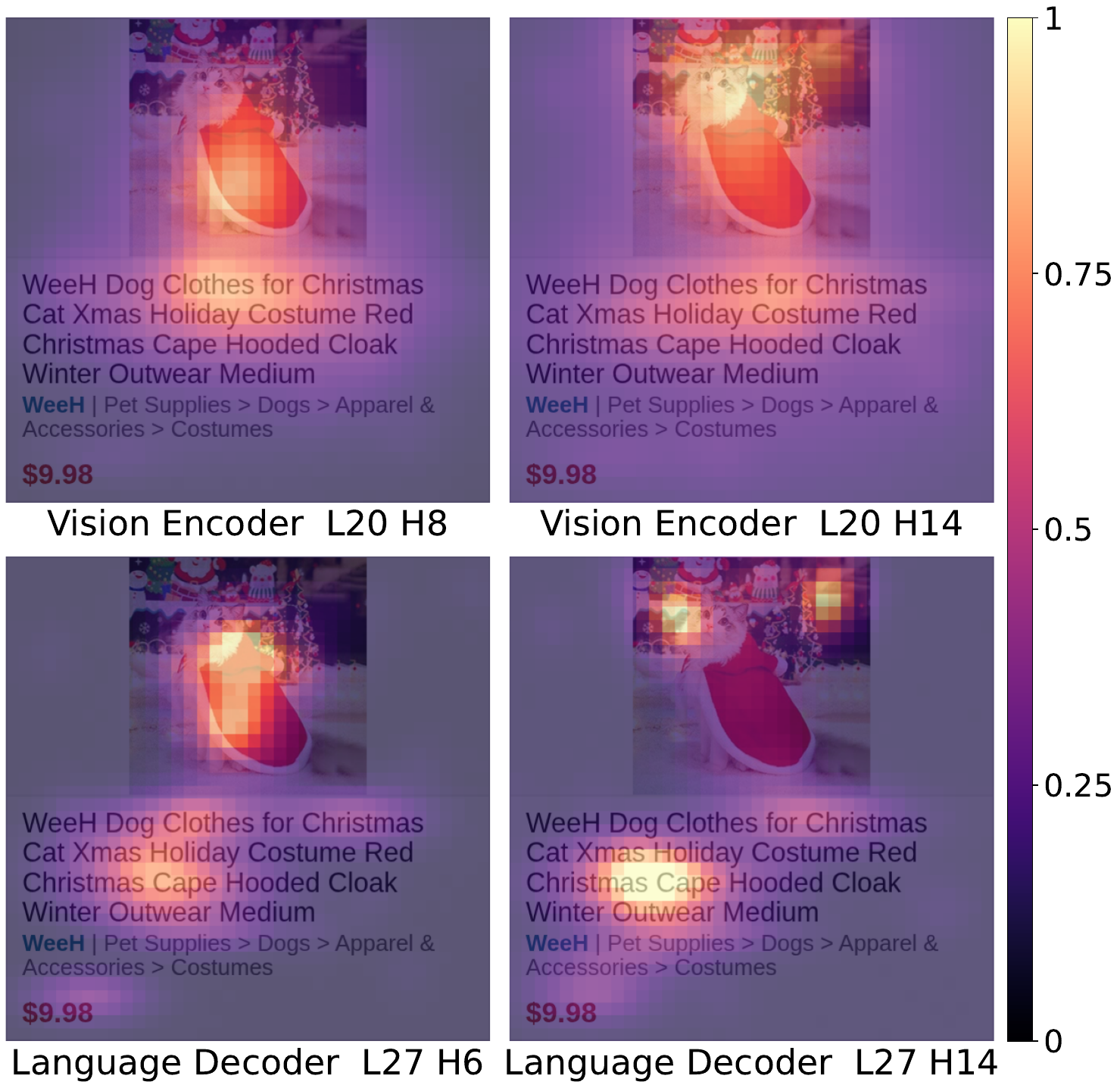}
    \caption{%
    Visualization of \ours's \textit{dual attention} mechanism on a Christmas cat-clothes item card, using the patches for “Cat,” “Christmas,” and “Clothes” as the query. L, H mean Layer, Head. The top and bottom rows show attention from the Qwen3-VL-8B vision encoder and language decoder, corresponding to the purple and green components in \Cref{fig: main figure}(c). Brighter colors indicate higher attention. Visual/Perception-level attention broadly contextualizes relevant regions, whereas language/semantic-level attention produces more concentrated image–text grounding.}
    \label{fig:dual attention visual}
\end{figure}
\subsection{Model and Embedding Behavior Analysis}
In this section, we investigate \ours's internal behavior by addressing three questions: What occurs during the encoding process? How do the resulting embeddings behave? Do they exhibit high representational quality?
\subsubsection{Dual Attention Mechanism of \ours's Encoding}
\label{subsec: Dual Attention Mechanism}
To further examine the introduced dual-attention mechanism during \ours's encoding process, we visualize the attention patterns of certain key information patches to all patches in the given item card. Specifically we demonstrate a Christmas cat-clothes item card and the query patches correspond to \textit{Cat}, \textit{Christmas}, and \textit{Clothes}. As shown in \Cref{fig:dual attention visual}. The vision encoder attends broadly to both the product image and related textual regions, suggesting perception-level
contextualization across visual objects and rendered glyphs. In contrast,
the language decoder exhibits more localized attention around semantically
matched image and text regions. These complementary patterns indicate that
\ours performs cross-modal interaction at two stages: broad perceptual
contextualization in the vision encoder, followed by more concentrated
semantic alignment in the language decoder.
\subsubsection{Embedding Cross-Modal Correlation Visualization}
\begin{figure}[t]
  \centering
  \includegraphics[width=\linewidth]{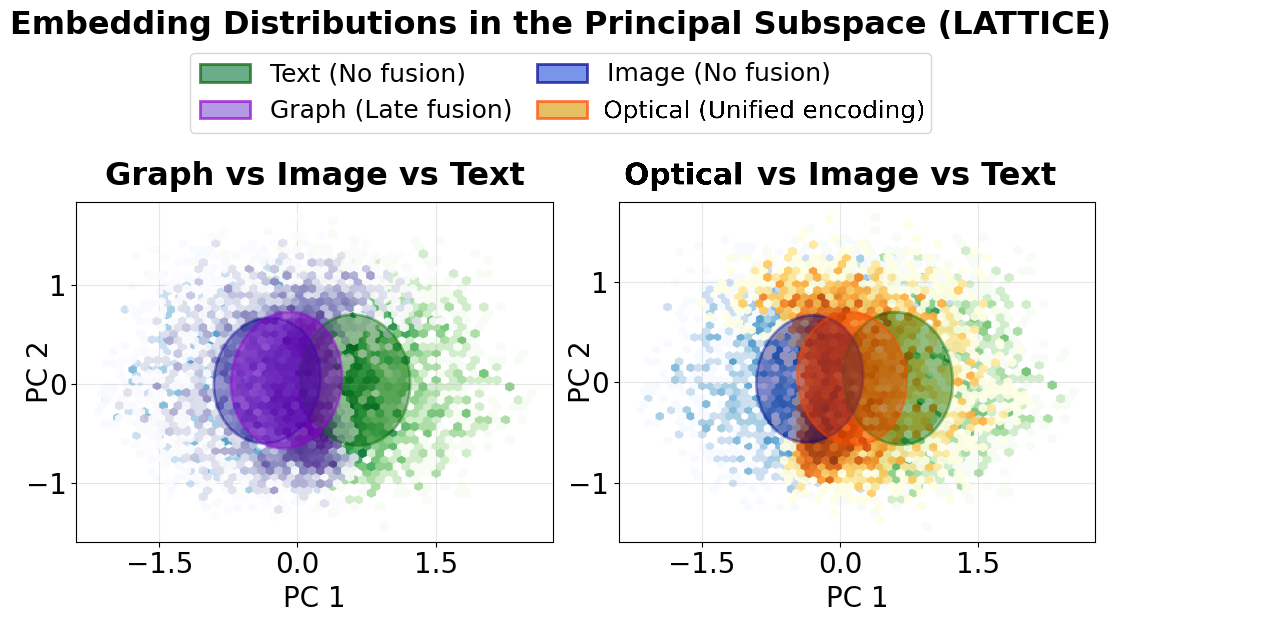}
  \caption{
  PCA visualization of embedding distributions with 50\% confidence ellipses under the graph-based backbone. \ours (Orange) forms a unified distribution bridging Image and Text representations, while Late Fusion (Purple) remains biased toward a single modality.
  }
  \label{fig:pca_singular_vectors}
\end{figure}
To further examine the embedding geometry behind effective rank, we visualize PCA-projected item representations of Image-only, Text-only, Late Fusion, and \ours using hexbin density plots with 50\% confidence ellipses. As shown in \Cref{fig:pca_singular_vectors}, \ours forms a unified manifold between image and text representations, suggesting stronger vision-language alignment. In contrast, Late Fusion is biased toward a single modality, i.e., Text in VBPR and Image in LATTICE, indicating that post-hoc fusion struggles to balance heterogeneous modality distributions.

\begin{figure}[t]
    \centering
    \includegraphics[width=\linewidth]{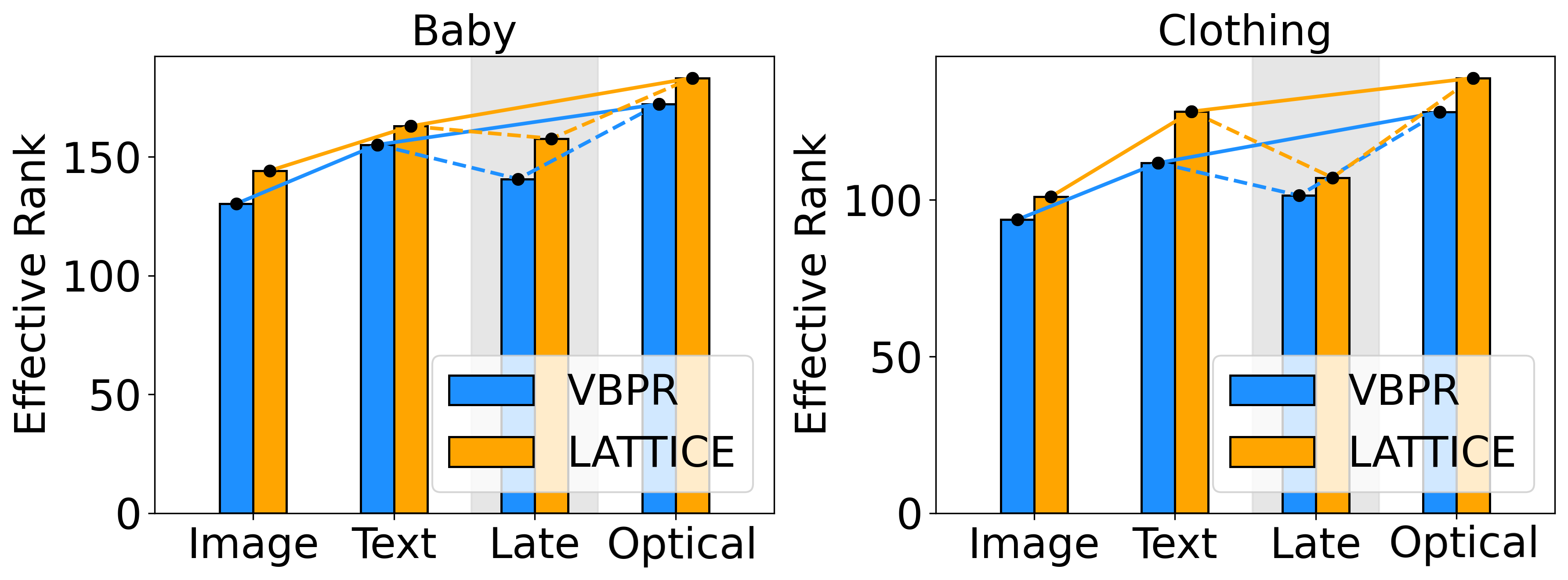}
    \caption{%
    Effective rank comparison among single-modality encoding, Late Fusion, and OpticalRec. Grey regions highlight the effective-rank drop caused by Late Fusion, while \ours consistently ranks highest.
    }
    \label{fig:erank_fusion}
\end{figure}
\subsection{Embedding Quality via Matrix Effective Rank}
\label{subsec: matrix effective rank results analysis}
We use matrix effective rank $\operatorname{erank}(\cdot)$ to assess the informativeness of item embeddings from image-only encoding, text-only encoding, late fusion, and \ours. Late fusion follows each backbone's default design: concatenation for VBPR and graph-based fusion for LATTICE. As shown in \Cref{fig:erank_fusion}:
\begin{enumerate}[label=\arabic*.,leftmargin=*]
    \item \ours produces the most informative embedding space, achieving the highest average effective rank of 155.6. This exceeds late fusion by 22.81\% and the strongest unimodal baseline, text-only encoding, by 11.54\%, indicating richer preservation of vision--language variations.
    \item Late fusion exhibits dimensional collapse, with an average effective rank of 126.7, 9.18\% below text-only encoding. This suggests that fusing independently encoded modalities may distort rather than enrich the embedding space.
\end{enumerate}
Due to space limit, we provide a Spearman correlation study of erank and model performance in Appendix~\ref{app: spearman correlation}, which shows high relevance.

%% file: materials/tab_main_v4.tex
\begin{table*}[t]
\centering
\caption{Performance comparison on three Amazon benchmarks. \textbf{Bold} = best per group per column; \underline{underline} = second best. \ours outperforms strong baseline encoding methods in most cases across datasets and models.}
\label{tab:main}
\setlength{\tabcolsep}{3.5pt}
\renewcommand{\arraystretch}{0.95}
\resizebox{0.86\textwidth}{!}{
\begin{tabular}{lccccccccc}
\toprule
\textbf{Model/Dataset}
  & \multicolumn{3}{c}{\textbf{Baby}}
  & \multicolumn{3}{c}{\textbf{Pets}}
  & \multicolumn{3}{c}{\textbf{Clothing}} \\

\midrule

\textbf{VBPR}
  & \textbf{R@20} & \textbf{N@20} & \textbf{H@20}
  & \textbf{R@20} & \textbf{N@20} & \textbf{H@20}
  & \textbf{R@20} & \textbf{N@20} & \textbf{H@20} \\
~~~+ Sum            & 0.0791 & 0.0348 & 0.0838 & 0.0981 & 0.0465 & 0.1072 & 0.0434 & 0.0201 & 0.0517 \\
~~~+ Concat         & 0.0791 & \underline{0.0355} & 0.0840 & \textbf{0.1000} & 0.0473 & 0.1091 & \underline{0.0454} & \underline{0.0212} & 0.0536 \\
~~~+ Prompt         & \underline{0.0796} & \underline{0.0355} & \underline{0.0843} & 0.0994 & \underline{0.0474} & \underline{0.1093} & 0.0453 & \underline{0.0212} & \underline{0.0540} \\
~~~+ \textbf{OpticalRec} & \textbf{0.0808} & \textbf{0.0359} & \textbf{0.0857} & \underline{0.0997} & \textbf{0.0477} & \textbf{0.1097} & \textbf{0.0460} & \textbf{0.0215} & \textbf{0.0551} \\
\midrule

\textbf{LATTICE}
  & \textbf{R@20} & \textbf{N@20} & \textbf{H@20}
  & \textbf{R@20} & \textbf{N@20} & \textbf{H@20}
  & \textbf{R@20} & \textbf{N@20} & \textbf{H@20} \\
~~~+ Sum            & \underline{0.0784} & \underline{0.0345} & \underline{0.0832} & 0.1036 & 0.0484 & 0.1137 & 0.0498 & \underline{0.0229} & 0.0579 \\
~~~+ Concat         & 0.0762 & 0.0332 & 0.0808 & 0.1026 & 0.0474 & 0.1121 & 0.0468 & 0.0219 & 0.0557 \\
~~~+ Graph          & 0.0770 & 0.0340 & 0.0815 & 0.1042 & \underline{0.0489} & 0.1135 & \textbf{0.0509} & 0.0224 & 0.0573 \\
~~~+ Prompt         & 0.0779 & 0.0338 & 0.0825 & \underline{0.1046} & 0.0484 & \underline{0.1142} & 0.0504 & \underline{0.0229} & \underline{0.0597} \\
~~~+ \textbf{OpticalRec} & \textbf{0.0803} & \textbf{0.0357} & \textbf{0.0849} & \textbf{0.1084} & \textbf{0.0513} & \textbf{0.1181} & \underline{0.0507} & \textbf{0.0230} & \textbf{0.0601} \\
\midrule

\textbf{FREEDOM}
  & \textbf{R@20} & \textbf{N@20} & \textbf{H@20}
  & \textbf{R@20} & \textbf{N@20} & \textbf{H@20}
  & \textbf{R@20} & \textbf{N@20} & \textbf{H@20} \\
~~~+ Sum            & 0.0820 & \underline{0.0358} & 0.0865 & \underline{0.0974} & \underline{0.0447} & \underline{0.1069} & 0.0416 & 0.0182 & 0.0493 \\
~~~+ Concat         & 0.0815 & \underline{0.0358} & 0.0862 & 0.0956 & 0.0428 & 0.1036 & 0.0402 & 0.0171 & 0.0481 \\
~~~+ Graph          & 0.0819 & 0.0356 & 0.0865 & 0.0957 & 0.0428 & 0.1037 & 0.0404 & 0.0173 & 0.0482 \\
~~~+ Prompt         & \underline{0.0825} & \underline{0.0358} & \underline{0.0870} & 0.0972 & 0.0439 & 0.1056 & \underline{0.0421} & \underline{0.0186} & \underline{0.0501} \\
~~~+ \textbf{OpticalRec} & \textbf{0.0834} & \textbf{0.0363} & \textbf{0.0882} & \textbf{0.1020} & \textbf{0.0473} & \textbf{0.1107} & \textbf{0.0429} & \textbf{0.0194} & \textbf{0.0510} \\
\midrule

\textbf{BM3}
  & \textbf{R@20} & \textbf{N@20} & \textbf{H@20}
  & \textbf{R@20} & \textbf{N@20} & \textbf{H@20}
  & \textbf{R@20} & \textbf{N@20} & \textbf{H@20} \\
~~~+ Sum            & 0.0765 & 0.0336 & 0.0812 & 0.0893 & 0.0399 & 0.0970 & 0.0380 & 0.0164 & 0.0452 \\
~~~+ Concat         & 0.0777 & 0.0341 & 0.0825 & 0.0876 & 0.0391 & 0.0951 & 0.0382 & 0.0165 & 0.0455 \\
~~~+ Supervised     & 0.0792 & 0.0347 & 0.0841 & \underline{0.0912} & \underline{0.0408} & \underline{0.0991} & 0.0389 & 0.0168 & 0.0463 \\
~~~+ Prompt         & \underline{0.0797} & \underline{0.0350} & \underline{0.0846} & 0.0906 & 0.0405 & 0.0984 & \underline{0.0392} & \underline{0.0169} & \underline{0.0467} \\
~~~+ \textbf{OpticalRec} & \textbf{0.0807} & \textbf{0.0354} & \textbf{0.0860} & \textbf{0.0947} & \textbf{0.0421} & \textbf{0.1024} & \textbf{0.0397} & \textbf{0.0171} & \textbf{0.0475} \\
\midrule

\textbf{LGMRec}
  & \textbf{R@20} & \textbf{N@20} & \textbf{H@20}
  & \textbf{R@20} & \textbf{N@20} & \textbf{H@20}
  & \textbf{R@20} & \textbf{N@20} & \textbf{H@20} \\
~~~+ Sum            & 0.0804 & 0.0353 & 0.0853 & 0.0944 & 0.0427 & 0.1023 & 0.0395 & 0.0170 & 0.0474 \\
~~~+ Concat         & 0.0797 & 0.0350 & 0.0846 & 0.0929 & 0.0420 & 0.1006 & 0.0391 & 0.0168 & 0.0469 \\
~~~+ Attentive      & 0.0818 & 0.0357 & 0.0868 & 0.0963 & 0.0435 & 0.1043 & \underline{0.0404} & \underline{0.0174} & \underline{0.0485} \\
~~~+ Prompt         & \underline{0.0823} & \underline{0.0359} & \underline{0.0876} & \underline{0.0972} & \underline{0.0439} & \underline{0.1053} & 0.0403 & 0.0173 & 0.0484 \\
~~~+ \textbf{OpticalRec} & \textbf{0.0835} & \textbf{0.0366} & \textbf{0.0887} & \textbf{0.1010} & \textbf{0.0454} & \textbf{0.1092} & \textbf{0.0412} & \textbf{0.0177} & \textbf{0.0491} \\
\bottomrule
\end{tabular}
}
\end{table*}

%% file: 6_ablation.tex
\section{Ablation, Robustness, and Efficiency Studies}
\label{sec: ablation}
In this section, we answer the questions of whether \ours is robust against rendering resolution, layout, rendered text font size and color, etc. We demonstrate with extensive experiments that the rendering operation of \ours is one-time, lightweight, and robust against above factors with minimal tuning effort.

\subsection{Rendering Resolution}
\begin{figure}[t!]
  \centering
  \includegraphics[width=\linewidth]{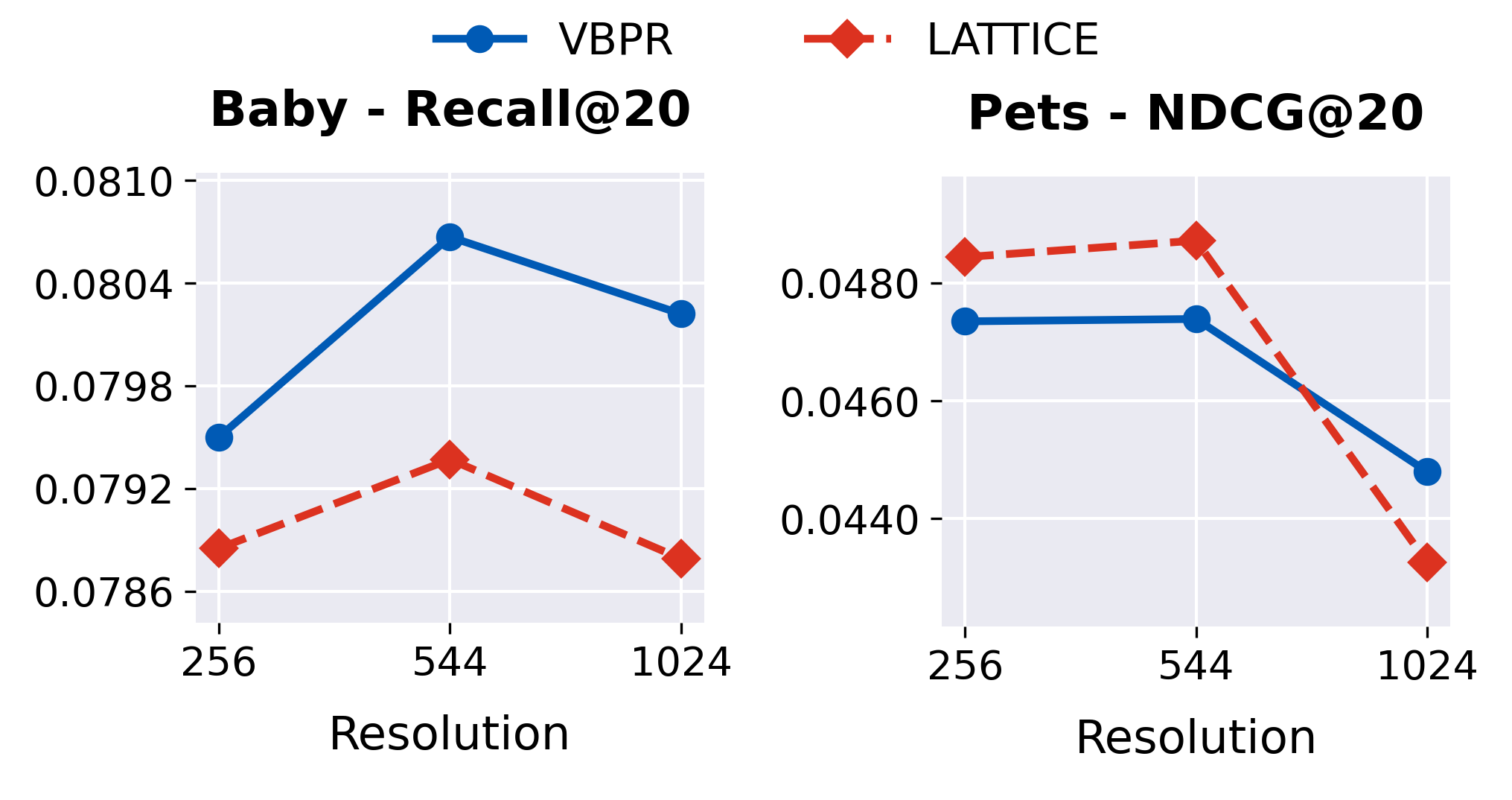}
  \caption{Performance under different Item Card resolutions. The $544 \times 544$ performs the best and is robust across datasets/models for efficient deployment.}
  \label{fig:resolution.png}
\end{figure}

We evaluate the impact of rendering resolution ($256 \times 256$), ($544 \times 544$), and ($1024 \times 1024$) pixel on the rendered Item Card. As shown in Figure~\ref{fig:resolution.png}, $544 \times 544$ achieves the best overall performance, while both lower and higher resolutions lead to slight degradation. This suggests that medium-resolution rendering already preserves sufficient visual-language semantics, avoiding unnecessary high-resolution vision-language-model inference and reducing resolution tuning cost.

\input{materials/tab_render_layout_font_size_color}
\subsection{Rendering Layout, Font Size and Color}
The item features we used for creating item cards are \{title, brand, category, price\}, which are naturally placed one-by-one below the item image as appeared similar on the amazon website. This follows most natural human visual perception without cherry picking order/coordinates. Additionally, we conducted a layout robustness study where we (1) randomly shuffle the position of \{title, brand category and price\} for 50\% of the data, (2) scale the font size by ±25\%, for 50\% data, and (3) change all rendered text color to black-and-white. The results are shown in \Cref{tab:layout_font_robustness}. We observe that \ours is robust against layout, font size and color by performing within ~±1\% for all controlled factors. We will also release the rendered cards and embeddings in the final version of the paper.

\input{materials/tab_longer_text}
\subsection{Rendering Longer Text}
We provide an experiment on the text trimming threshold of the item's detailed textual description in \Cref{tab:text_length}. We observe that using longer text (detailed item description) does not require heavy resolution scale down nor VLM to degrade but instead improves +0.7\% to +1.6\% performance. Statistically, a maximum text length of 1500 char includes ~94\% of items in our dataset, and could be properly rendered onto 512*1024 item cards for both human and machine reading with optical character recognition (OCR) rate close to ~100\%. We find the longer text benefits saturates at the 1000 char threshold (+1.14\% improvement). This also suggests that our default setup without item detailed description, but rather {title, brand, category, price} already provides a strong recommendation signal for user-item matching.

\subsection{Rendering Efficiency}
\input{materials/tab_render_efficiency}
We report the one-time offline rendering overhead in \Cref{tab:rendering_efficiency}. Even for the largest Clothing dataset, rendering 75,616 items takes 57 minutes with 8 workers and 10.1 GB storage, indicating practical overhead on modern machines. Together with the resolution robustness in \Cref{fig:resolution.png}, \ours does not require repeated high-resolution rendering.

\section{Conclusions}
We present \ours, a unified optical vision-language encoding framework for multimodal collaborative filtering. By rendering item text and images into a shared visual space, \ours enables complementary perceptual- and semantic-level cross-modal interactions while avoiding disruptive late fusion. Extensive experiments demonstrate consistent gains across datasets and recommendation backbones, together with improved embedding quality, robustness to rendering variations, and practical one-time preprocessing cost. These results highlight unified optical encoding as a simple and effective alternative to conventional independently encoded multimodal representations.

%% file: materials/tab_render_layout_font_size_color.tex
\begin{table}[t]
\centering
\caption{Robustness against layout, font scale and color variations. All results (Recall@20) remain within $\pm1\%$ relative deviation from the default setting.}
\small
\setlength{\tabcolsep}{5pt}
\renewcommand{\arraystretch}{0.92}
\begin{tabular}{llccc}
\toprule
\textbf{Model} & \textbf{Variant}
& \textbf{Baby}
& \textbf{Pets}
& \textbf{Cloth} \\
\midrule
\multirow{3}{*}{LATTICE}
& Ours       & 0.0803 & 0.1084 & 0.0506 \\
& Layout     & 0.0798 & 0.1078 & 0.0509 \\
& Font Scale & 0.0799 & 0.1092 & 0.0509 \\
& Font Color & 0.0802 & 0.1080 & 0.0504 \\
\midrule
\multirow{3}{*}{FREEDOM}
& Ours       & 0.0834 & 0.1020 & 0.0429 \\
& Layout     & 0.0829 & 0.1025 & 0.0427 \\
& Font Scale & 0.0833 & 0.1014 & 0.0432 \\
& Font Color & 0.0830 & 0.1012 & 0.0430 \\
\midrule
\multirow{3}{*}{LGMRec}
& Ours       & 0.0835 & 0.1010 & 0.0412 \\
& Layout     & 0.0834 & 0.1011 & 0.0411 \\
& Font Scale & 0.0837 & 0.1012 & 0.0409 \\
& Font Color & 0.0833 & 0.1009 & 0.0408 \\
\bottomrule
\end{tabular}
\label{tab:layout_font_robustness}
\end{table}

%% file: materials/tab_longer_text.tex
\begin{table}[tb]
\centering
\caption{Performance for various maximum text-length thresholds. Best result for each backbone is in bold.}
\label{tab:text_length}
\resizebox{\columnwidth}{!}{%
\begin{tabular}{llcccccc}
\toprule
\multirow{2}{*}{\textbf{}} 
& \multirow{2}{*}{\textbf{Variant}} 
& \multicolumn{2}{c}{\textbf{Baby}} 
& \multicolumn{2}{c}{\textbf{Pets}} 
& \multicolumn{2}{c}{\textbf{Cloth}} \\
\cmidrule(lr){3-4} 
\cmidrule(lr){5-6} 
\cmidrule(lr){7-8}
& 
& \textbf{R@20} & \textbf{N@20} 
& \textbf{R@20} & \textbf{N@20} 
& \textbf{R@20} & \textbf{N@20} \\
\midrule

\multirow{4}{*}{\rotatebox[origin=c]{90}{\textbf{\small LATTICE}}}
& Ours 
& 0.0803 & 0.0357 
& 0.1084 & 0.0513 
& 0.0507 & 0.0230 \\
& $<500$ char 
& \textbf{0.0813} & \textbf{0.0362} 
& 0.1095 & \textbf{0.0520} 
& 0.0511 & 0.0232 \\
& $<1000$ char 
& 0.0812 & 0.0361 
& \textbf{0.1096} & 0.0519 
& \textbf{0.0514} & \textbf{0.0233} \\
& $<1500$ char 
& 0.0809 & 0.0360 
& \textbf{0.1096} & \textbf{0.0520} 
& 0.0512 & \textbf{0.0233} \\
\midrule

\multirow{4}{*}{\rotatebox[origin=c]{90}{\textbf{\small FREEDOM}}}
& Ours 
& 0.0834 & 0.0363 
& 0.1020 & 0.0473 
& 0.0429 & 0.0194 \\
& $<500$ char 
& \textbf{0.0845} & \textbf{0.0368} 
& 0.1030 & 0.0479 
& \textbf{0.0433} & \textbf{0.0196} \\
& $<1000$ char 
& 0.0844 & \textbf{0.0368} 
& 0.1033 & 0.0480 
& 0.0432 & \textbf{0.0196} \\
& $<1500$ char 
& 0.0840 & 0.0366 
& \textbf{0.1036} & \textbf{0.0482} 
& 0.0432 & \textbf{0.0196} \\
\midrule

\multirow{4}{*}{\rotatebox[origin=c]{90}{\textbf{\small LGMRec}}}
& Ours 
& 0.0835 & 0.0366 
& 0.1010 & 0.0454 
& 0.0412 & 0.0177 \\
& $<500$ char 
& 0.0844 & 0.0370 
& 0.1017 & 0.0458 
& 0.0416 & 0.0179 \\
& $<1000$ char 
& \textbf{0.0845} & \textbf{0.0371} 
& 0.1018 & 0.0459 
& \textbf{0.0418} & \textbf{0.0180} \\
& $<1500$ char 
& 0.0841 & 0.0369 
& \textbf{0.1025} & \textbf{0.0462} 
& 0.0416 & 0.0179 \\
\bottomrule
\end{tabular}%
}
\end{table}

%% file: materials/tab_render_efficiency.tex
\begin{table}[t!]
\centering
\caption{Rendering efficiency as one-time offline process. Speed means processed cards per second.}
\label{tab:rendering_efficiency}
\resizebox{\columnwidth}{!}{%
\begin{tabular}{lrrrrr}
\toprule
\textbf{Dataset} & \textbf{Items} & \textbf{Speed} & \textbf{Workers} & \textbf{Minutes} & \textbf{Storage (GB)} \\
\midrule
Baby     & 17,836 & 22 & 8 & 13.5 & 2.4 \\
Pets     & 61,060 & 22 & 8 & 46.0 & 8.1 \\
Clothing & 75,616 & 22 & 8 & 57.0 & 10.1 \\
\bottomrule
\end{tabular}%
}
\end{table}

%% file: 7_appendix.tex
\appendix

\section{Mutual Information Gap and Practical Implications}
\subsection{Mutual Information Gap Proof}
\label{app:mutual_information_gap}
Equation (7) compares the expressiveness between \ours and late-fusion representations about the original image-text input by defining the information gap. We here provide a more detailed proof of why this gap is strictly positive.

\paragraph{The cross-modal term $\mathcal{I}_{\text{cross}}^{\text{OR}} > 0$.}
Let $\mathbf{x}_i=(\mathbf{v}_i,\mathbf{t}_i)$ denotes the original image-text input. Remember $\mathbf{s}_i^{vt}$ denotes the vision-language layout structure defined in \Cref{subsec: multimodal cf definition}. We define the cross-modal gain as the extra information preserved by the unified representation over the late-fusion intermediate embedding:
\begin{equation}
  \mathcal{I}_{\text{cross}}^{\text{OR}}
  \triangleq
  I(\mathbf{q}_i^{\text{OR}};\mathbf{x}_i)
  -
  I(\mathbf{z}_i^{\text{LF}};\mathbf{x}_i).
\end{equation}
Let $\mathbf{c}_i$ denote cross-modal information contained in $\mathbf{s}_i^{vt}$, Since late fusion independently encodes $\mathbf{v}_i$ and $\mathbf{t}_i$, such information is not fully recoverable from $\mathbf{z}_i^{\text{LF}}$:
\begin{equation}
  I(\mathbf{c}_i;\mathbf{x}_i\mid \mathbf{z}_i^{\text{LF}})>0.
\end{equation}
In contrast, \ours exposes $\mathbf{c}_i$ implied by $\mathbf{s}_i^{vt}$ to a single vision-language
encoder through unified pixel-space rendering. For the resulting non-degenerate cross-modal channel, $I(\mathbf{q}_i^{\text{OR}};\mathbf{c}_i\mid\mathbf{z}_i^{\text{LF}})>0$, and thus $\mathcal{I}_{\text{cross}}^{\text{OR}}>0$.
 
\paragraph{The distortion term $\mathcal{D}_{\text{intra}}^{\text{LF}} > 0$.}
Formally the distortion term decomposes into:
\begin{equation}
  \mathcal{D}_{\text{intra}}^{\text{LF}}
  \triangleq
  I(\mathbf{z}_i^{\text{LF}};\mathbf{x}_i)
  -
  I(\mathbf{q}_i^{\text{LF}};\mathbf{x}_i).
\end{equation}
where the right-hand side denotes information retained before fusion subtracts that retained after fusion. Since late fusion is defined as $\mathbf{q}_i^{\text{LF}}=g(\mathbf{z}_i^{\text{LF}})$, we have the
Markov chain
$\mathbf{x}_i \rightarrow \mathbf{z}_i^{\text{LF}} \rightarrow
\mathbf{q}_i^{\text{LF}}$. By the data processing inequality
(DPI)~\cite{cover2006}:
\begin{equation}
  I(\mathbf{q}_i^{\text{LF}};\mathbf{x}_i)
  \leq
  I(\mathbf{z}_i^{\text{LF}};\mathbf{x}_i),
\end{equation}
and thus $\mathcal{D}_{\text{intra}}^{\text{LF}}\geq 0$. Equality would require late fusion $g(\cdot)$ be strictly lossless compression, which does not hold for our multimodal recommendation settings. Hence $\mathcal{D}_{\text{intra}}^{\text{LF}}>0$ for
standard lossy late fusion.

\subsection{Information Gap Practical Implications}
\label{app: Information Gap Practical Implications}
While the mutual information terms in \Cref{eq:gap} are difficult to
estimate directly for high-dimensional LVLM representations, unified
encoding enables a large cross-modal interaction budget. Specifically,
image patch tokens and text glyph tokens are processed in the same
attention sequence, giving
\begin{equation}
  \mathcal{B}_{\text{cross}}^{\text{OR}}
  \approx
  L \cdot H \cdot |\mathcal{V}| \cdot |\mathcal{T}| \cdot d_h,
  \label{eq:activation}
\end{equation}
where $L$, $H$, and $d_h$ are the number of layers, heads, and per-head
dimension. For Qwen3-VL-8B with $L{=}32$, $H{=}32$, $d_h{=}128$, and
$|\mathcal{V}|\approx|\mathcal{T}|\approx256$, this yields
$\mathcal{B}_{\text{cross}}^{\text{OR}}\approx8.6\times10^{9}$ potential
cross-modal interactions. This suggests
that the gap in \Cref{eq:gap} is practically substantial regarding representation capacity. To further accurately quantify the information gap and data variances in \ours and Late-fusion encoding, we provide one more theoretical analysis in \Cref{subsec: method matrix effective rank} via matrix effective-rank $erank(\cdot)$.


\subsection{Detailed Derivation of Effective Rank}
\label{app:effective_rank_derivation}

Different from the matrix algebraic/true rank, the effective rank $erank(\cdot)$ down-weights near-zero singular values and therefore better reflects representation concentration or collapse~\cite{roy2007effective, Su2025effective}. Consider a real-valued matrix $\mathbf{X} \in \mathbb{R}^{m \times n}$ which is the item embedding matrix/multimodal feature matrix in our setting, we first conduct singular value decomposition of $\mathbf{X}$:

\begin{equation}
\resizebox{0.87\linewidth}{!}{$
\mathbf{X} = \mathbf{U}\,\boldsymbol{\Sigma}\,\mathbf{V}^{\top},
\quad
\mathbf{U} \in \mathbb{R}^{m \times r},\ 
\boldsymbol{\Sigma} \in \mathbb{R}^{r \times r},\ 
\mathbf{V} \in \mathbb{R}^{n \times r}
$}
\end{equation}

We assume $n=r$ for simplicity, then $n = r = \operatorname{rank}(\mathbf{X})$. Let $\boldsymbol{\Sigma} = \operatorname{diag}(\sigma_1, \dots, \sigma_n)$ denote
the diagonal matrix of singular values, where $\sigma_i \geq 0$, by normalization we can further construct a probability distribution over all singular values as

\begin{equation}
p_i = \frac{\sigma_i}{\sum_{j=1}^n \sigma_j}
\label{eq: sigma i to p i}
\end{equation}

Based on the constructed probability $p_i$, we can further compute the Shannon entropy as

\begin{equation}
\text{H(p)} = -\sum_{i=1}^n p_i \log p_i
\end{equation}

$\operatorname{erank}(\mathbf{X})$ can be further defined as the exponential of Shannon entropy, satisfying one desired math property of $\mathop{\text{erank}}(\mathbf{X})$ range discussed later:

\begin{equation}
\mathop{\text{erank}}(\mathbf{X}) = \exp\left(-\sum_{i=1}^n p_i \log p_i\right)
\label{eq: erank of p i}
\end{equation}

As mentioned earlier and introduced in~\cite{roy2007effective}, the above definition satisfies one desired math property of $\mathop{\text{erank}}(\mathbf{X})$ range:

\begin{property}
\label{property:erank-extremes}
Let $\mathbf{p} = (p_1, \dots, p_n)$ be a probability distribution induced by
the singular values, and let $\mathrm{erank}(\mathbf{X}) = \exp(H(\mathbf{p}))$, where $\text{H}(\mathbf{p}) = - \sum_{i=1}^n p_i \log p_i$, then the effective rank satisfies:
\[
1 \le \mathrm{erank}(\mathbf{X}) \le n .
\]
where $\mathrm{erank}(\mathbf{X}) = 1$ if and only if $\mathbf{p}$ follows a one-hot distribution; and $\mathrm{erank}(\mathbf{X}) = n$ if $\mathbf{p}$ follows a uniform distribution.
\end{property}

\section{Dataset, Baseline, and Implementation Details}
\subsection{Dataset Details}
\label{app: dataset details}
For evaluating our models and fair comparison, we select three most commonly used benchmarks by multimodal collaborative filtering recommendation from \emph{Amazon.com} known for real-world sparsity, namely \emph{Baby Products} (Baby), \emph{Pet Supplies} (Pets), and \emph{Clothing, Shoes \& Jewelry} (Clothing)~\cite{hou2024bridging}. For preprocessing, we follow~\cite{malitesta2025formalizing, pomo2025recommender} to apply 5-core filtering to exclude users and items with less than five-time appearances. For textural feature selection, we choose \emph{title}, \emph{price}, \emph{brand} and \emph{categories}; For visual features, we use \emph{photos} of the items. We also filter out items without above metadata. Detailed statistics of the datasets are reported in \Cref{tab:dataset_stats} including users (\#User), items (\#Item), interactions (\#Inter.) and dataset sparsity in percentages.

\subsection{Baseline Methods}
\label{app: baseline methods}
\begin{itemize}[label=\small$\bullet$,leftmargin=*]
\item\textbf{VBPR}~\cite{he2016vbpr} is an early multimodal collaborative filtering method that extends BPR with CNN-based visual features to model visually driven user preferences.
\item\textbf{LATTICE}~\cite{zhang2021mining} constructs modality-aware item-item similarity graphs for each modality and applies graph convolutions to inject high-order item affinities into item representations for downstream collaborative filtering.
\item\textbf{FREEDOM}~\cite{zhou2023tale} freezes the item-item graph and applies degree-sensitive edge pruning to denoise the user-item interaction graph, improving robustness over LATTICE's dynamic graph learning.
\item\textbf{BM3}~\cite{zhou2023bootstrap} proposes a self-supervised multimodal framework that uses dropout-based contrastive views, eliminating the need for negative sampling and auxiliary graph augmentations.
\item\textbf{LGMRec}~\cite{guo2024lgmrec} captures both local and global modality correlations via hypergraph structures, modeling fine-grained cross-modal interactions for multimodal preference alignment. We additionally construct a strong prompt-based VLM baseline (Figure~1(b)) as a controlled comparison, isolating the benefit of OpticalRec's additional perception-level image--glyph interaction beyond decoder-level semantic interaction.
\end{itemize}

\subsection{Implementations}
\label{app: Implementations}
We adopt Qwen3-VL-8B for main results report. All vision-language models are frozen during encoding process for computational efficiency and for comparing with previous works. For training \ours and all baseline recommendation models, we utilize AdamW optimizer with learning rate of 1e-3/1e-4 with maximum epochs of 500. Validation is performed per epoch and the training is stopped once validation performance does not improve for 10 epochs. The model with best validation performance is saved for testing and metrics report. For hyperparameters, we find (1) embedding/model size, the number of Graph Neural Network layers, dropout rate and weight decay be the most sensitive ones for model performance. Specifically, we grid-search the embedding/model size in [64, 128, 256, 512, 1024, 2048], the number of Graph Neural Network layers in [1,2,4,8], dropout rate from [0.1,...,0.8] on a 0.1-stride and weight decay from [1e-6, 1e-4, 1e-2]. For \ours rendering operation, we tune the resolution among [256, 544, 1024], we analyze the efficiency of rendering and resolution tuning process in \Cref{sec: ablation}. For validation and test, we adopt three most commonly utilized metrics~\cite{malitesta2025formalizing, pomo2025recommender} as NDCG@$20$ and Recall@$20$ and H@$20$ (Hit Ratio), typical for recommendation algorithm evaluation.

\section{Additional Results}
\input{materials/tab_intern_vl}
\subsection{Ablation using InternVL3.5 as backbone VLM}
\label{app: internvl results}
We report the results for another SoTA VLM model as InternVL3.5-8B in \Cref{tab:internvl results} to highlight the plug-and-play benefit of OpticalRec. OpticalRec performs the best across all datasets and the average gain over the second best baseline is +2.51\%, the theoretical proof remains the same as it is model-irrelevant.

\subsection{Spearman Correlation Between Effective Rank and Model Performance}
\label{app: spearman correlation}
\input{materials/tab_spearman_correlation}
We here provide an Spearman correlation study of erank and model performance in \Cref{tab:spearman_correlation}. Overall, the Spearman rank correlation between effective rank and R@20 is $\rho = 0.85 (p < 0.001)$, indicating a strong and statistically significant positive relationship.

%% file: materials/tab_intern_vl.tex
\begin{table}[tb]
\centering
\caption{Comparison of late fusion, prompt-based encoding, and \ours using InternVL3.5-8B as backbone VLM. The best result within each recommendation model is highlighted in bold.}
\label{tab:internvl results}
\resizebox{\columnwidth}{!}{%
\begin{tabular}{llcccccc}
\toprule
\multirow{2}{*}{\textbf{}} 
& \multirow{2}{*}{\textbf{Variant}} 
& \multicolumn{2}{c}{\textbf{Baby}} 
& \multicolumn{2}{c}{\textbf{Pets}} 
& \multicolumn{2}{c}{\textbf{Cloth}} \\
\cmidrule(lr){3-4}
\cmidrule(lr){5-6}
\cmidrule(lr){7-8}
&
& \textbf{R@20} & \textbf{N@20}
& \textbf{R@20} & \textbf{N@20}
& \textbf{R@20} & \textbf{N@20} \\
\midrule

\multirow{3}{*}{\rotatebox{90}{\resizebox{3.8em}{!}{\textbf{LATTICE}}}}
& + Late
& 0.0771 & 0.0342
& 0.1018 & 0.0474
& 0.0487 & 0.0221 \\
& + Prompt
& 0.0775 & 0.0345
& 0.1036 & 0.0484
& 0.0494 & 0.0225 \\
& + \ours
& \textbf{0.0794} & \textbf{0.0352}
& \textbf{0.1075} & \textbf{0.0503}
& \textbf{0.0501} & \textbf{0.0227} \\
\midrule

\multirow{3}{*}{\rotatebox{90}{\resizebox{3.8em}{!}{\textbf{FREEDOM}}}}
& + Late
& 0.0796 & 0.0351
& 0.0941 & 0.0435
& 0.0402 & 0.0181 \\
& + Prompt
& 0.0804 & 0.0354
& 0.0974 & 0.0448
& 0.0416 & 0.0186 \\
& + \ours
& \textbf{0.0817} & \textbf{0.0359}
& \textbf{0.1004} & \textbf{0.0463}
& \textbf{0.0421} & \textbf{0.0191} \\
\midrule

\multirow{3}{*}{\rotatebox{90}{\resizebox{3.8em}{!}{\textbf{LGMRec}}}}
& + Late
& 0.0795 & 0.0346
& 0.0935 & 0.0418
& 0.0390 & 0.0169 \\
& + Prompt
& 0.0805 & 0.0348
& 0.0963 & 0.0434
& 0.0399 & 0.0171 \\
& + \ours
& \textbf{0.0825} & \textbf{0.0359}
& \textbf{0.0997} & \textbf{0.0448}
& \textbf{0.0407} & \textbf{0.0176} \\
\bottomrule
\end{tabular}%
}
\end{table}

%% file: materials/tab_spearman_correlation.tex
\begin{table}[tb]
\centering
\caption{Spearman correlation between Recall@20 and matrix effective rank.}
\label{tab:spearman_correlation}
\resizebox{\columnwidth}{!}{%
\begin{tabular}{llccc}
\toprule
\textbf{Dataset} & \textbf{Model} & $\boldsymbol{\rho}$ & \textbf{$p$-value} & \textbf{Relevancy} \\
\midrule
Baby     & VBPR    & +1.0000 & $<0.05$ & Most relevant \\
Baby     & LATTICE & +0.8000 & 0.2000  & Highly relevant \\
Clothing & VBPR    & +0.8000 & 0.2000  & Highly relevant \\
Clothing & LATTICE & +0.4000 & 0.6000  & Medium relevant \\
\midrule
\textbf{Overall} & --- & \textbf{+0.8471} & \textbf{$<0.001$} & \textbf{Highly relevant} \\
\bottomrule
\end{tabular}%
}
\end{table}